\documentclass[conference]{IEEEtran}

\usepackage{bm}
\usepackage{amsfonts}
\usepackage{amsmath}
\usepackage{amssymb}
\usepackage{lipsum}

\usepackage{multirow}
\usepackage{booktabs}
\usepackage{array}

\usepackage[noadjust]{cite}

\usepackage{graphicx}
\graphicspath{{./figure/}}

\usepackage{pgfplots}
\pgfplotsset{compat=1.15}

\usepackage[hidelinks]{hyperref}

\usepackage{array}
\newcommand{\PreserveBackslash}[1]{\let\temp=\\#1\let\\=\temp}
\newcolumntype{C}[1]{>{\PreserveBackslash\centering}p{#1}}
\newcolumntype{R}[1]{>{\PreserveBackslash\raggedleft}p{#1}}
\newcolumntype{L}[1]{>{\PreserveBackslash\raggedright}p{#1}}

\DeclareMathOperator*{\diag}{diag}
\DeclareMathOperator*{\argmin}{arg\,min}

\newcommand{\R}{\mathbb{R}}

\newcommand{\bO}{\mathcal{O}}
\newcommand{\G}{\mathcal{G}}
\newcommand{\V}{\mathcal{V}}
\newcommand{\E}{\mathcal{E}}

\title{Cyberattack Detection in Large-Scale Smart Grids using Chebyshev Graph Convolutional Networks}

\date{\today}

\begin{document}

\author{
	\IEEEauthorblockN{Osman Boyaci}
	\IEEEauthorblockA{
		Electrical Engineering\\
		Texas A\&M University\\
		College Station, TX, 77843\\
		osman.boyaci@tamu.edu
	}  \and
	\IEEEauthorblockN{M. Rasoul Narimani}
	\IEEEauthorblockA{
		College of Engineering\\
		Arkansas State University\\
		Jonesboro, AR, 72404\\
		mnarimani@astate.edu
	} \and
	\IEEEauthorblockN{Katherine Davis}
	\IEEEauthorblockA{
		Electrical Engineering\\
		Texas A\&M University\\
		College Station, TX, 77843\\
		katedavis@tamu.edu
	} \and
	\IEEEauthorblockN{Erchin Serpedin}
	\IEEEauthorblockA{
		Electrical Engineering\\
		Texas A\&M University\\
		College Station, TX, 77843\\
		eserpedin@tamu.edu
	}
}


\maketitle

\begin{abstract}
As a highly complex and integrated cyber-physical system, modern power grids are exposed to cyberattacks. False data injection attacks (FDIAs), specifically, represent a major class of cyber threats to smart grids by targeting the measurement data's integrity. Although various solutions have been proposed to detect those cyberattacks, the vast majority of the works have ignored the inherent graph structure of the power grid measurements and validated their detectors only for small test systems with less than a few hundred buses. To better exploit the spatial correlations of smart grid measurements, this paper proposes a deep learning model for cyberattack detection in large-scale AC power grids using Chebyshev Graph Convolutional Networks (CGCN). By reducing the complexity of spectral graph filters and making them localized, CGCN provides a fast and efficient convolution operation to model the graph structural smart grid data. We numerically verify that the proposed CGCN based detector surpasses the state-of-the-art model by 7.86\% in detection rate and 9.67\% in false alarm rate for a large-scale power grid with 2848 buses. It is notable that the proposed approach detects cyberattacks under 4 milliseconds for a 2848-bus system, which makes it a good candidate for real-time detection of cyberattacks in large systems.
\end{abstract}

\section{Introduction}
A modern power grid integrates the cyber communication network into the physical power system infrastructure.
In this highly complex cyber-physical system, Remote Terminal Units (RTUs) deliver physical measurement data to the Supervisory Control and Data Acquisition Systems (SCADAs) \cite{davis2012power}.
Then, the communication network transfers these measurements to the application level in which they are processed by the Energy Management System (EMS). The security of the cyber-physical pipeline is critical as the secure and reliable operation of power grids strongly depends on the integrity of these data.

Integrity and validity of input data for the power system state estimation (PSSE) algorithm is crucial for the reliability of power grid operations as the output of the PSSE block is directly used by various EMS units such as forecasting of the load and analysis of contingency \cite{giannakis2013monitoring}. As one of the major classes of cyberattacks to the PSSE, false data injection attacks (FDIAs) aim to compromise the measurement data to bypass the bad data detection (BDD) algorithm and make the attack unobservable
\cite{liang2016review}. 
If the grid operator takes actions according to the false system state, s/he might jeopardize the security and reliability of the grid. Traditional BDD algorithms for detecting FDIAs are insufficient as stealth (unobservable) cyberattacks can easily dodge these algorithms.
Thus, FDIAs constitutes one of the most prominent threats to today's modern power grids.

FDIA detection algorithms are classified into two main categories as model-based methods and data-driven methods \cite{musleh2019survey}.
In model-based methods, a system model is built and its parameters are estimated. They do not require a historical dataset since they do not include any separate system to be trained.  
However, manual threshold optimization steps, high detection delays, and scalability issues limit their applicability for real time analysis.
In contrast, data-driven methods eliminate the manual tuning steps, increase the scalability of the algorithm for attack detection, and reduce the detection time at the expense of a training process which need a historical dataset \cite{musleh2019survey}.

Recently, deep learning (DL) based data-driven detectors such as Fully-Connected Neural Networks (FCN) \cite{s97}, Recurrent Neural Network (RNN) \cite{s94}, Convolutional Neural Network (CNN) \cite{lu2020false} have been proposed for cyberattack detection in power grids, thanks to the increasing volume of collected historical data samples.
However, despite their powerful modeling capabilities, DL approaches may not generalize the data well enough and fail to detect cyberattacks if their architecture ignores the underlying physical system generating the data \cite{musleh2019survey}. 
For instance, RNNs are perfectly suited architectures to model the recurrent structure of the language data.
Similarly, CNNs are better candidates for image and video processing since sliding kernels are extremely efficient to exploit the pixel locality of image data \cite{deep_learning}. 

Most of the works dealing with the detection of FDIAs ignore the spatial correlation of the power grid data, therefore, they fail to fully model the underlying graph topology of the power grid.
There is only a few works \cite{drayer2019detection, ramakrishna2019detection} in the literature that exploit the spatial correlations of the power grids to detect cyberattacks using Graph Signal Processing (GSP).
Although GSP offers highly efficient tools for cyberattack detection, the custom design steps of spectral filters and scalability problems restrict its usability in real life scenarios. 
As a data-driven version of GSP, Graph Convolutional Networks (GCN) predict the filter weights in their hidden layers and eliminate the manual filter design steps.
This automation makes GCNs more attractive to power grid applications, for instance, GCNs are utilized for optimal power flow applications of power grids in \cite{owerko2020optimal}, for cyberattack detection in \cite{boyaci2021graph}, and for cyberattack localization in \cite{boyaci2021joint}. 

Scalability and detection time are the most fundamental factors in designing cyberattack detectors as the number of units varies between tens to a few thousand in today's power grids.
Except a few highly scalable models including \cite{deng2015defending, liu2014detecting, boyaci2021graph,boyaci2021joint}, most of the proposed models for cyberattack detection are designed for small-scale systems such as IEEE 14- \cite{s45, s60, s65, s85} and IEEE 30- \cite{s95, s94} bus test systems.  
Scalability problems may appear as high detection delays when small-scale models are applied to large-scale networks.

In this work, we propose a cyberattack detection model that utilizes Chebsyhev Graph Convolutional Networks in its hidden layers, enabling it to fully exploit the spatial correlations of the smart grid data.
We integrate the grid topology to our detector with the weighted graph adjacency matrix obtained by grid's admittance matrix represented with $\bm{Ybus}$.
To train the proposed model, we generate a historical dataset having 36000 samples for a large-scale smart grid with 2848 buses.
We verify the proposed detector by implementing two of the frequently used cyberattack models and comparing the detection results with the existing architectures designed for cyberattack detection.

The contributions of this work are summarized as follows:
\begin{itemize}
	\item We design a deep learning architecture by employing Chebyshev Graph Convolutional Networks in its hidden layers to adequately capture the spatial correlations of graph structural smart grid data.
	\item We propose a model to detect cyberattacks in a few milliseconds even for large-scale grids with more than 2000 buses.
	\item The proposed model has an end-to-end automatic training process without having any custom optimization step.
\end{itemize}

The rest of this paper is organized as follows.
While Section~\ref{sec:problem} formulates the problem,
Section~\ref{sec:methods} proposes the FIDA detection method using GCNN.
Results and discussion are presented in Section~\ref{sec:results}.
Section~\ref{sec:conclusion} concludes the paper. 

\section{Problem Formulation}\label{sec:problem}
In power grids, system state $\bm{x} \in \R^n$ is calculated in PSSE block using complex measurements $\bm{z} \in \R^m$ as follows:
\begin{equation} \label{eq:psse}
\hat{\bm{x}} = \argmin\limits_x || \bm{z} - h(\bm{x})||^2.
\end{equation}
where $\bm{x}$ represents bus voltage magnitudes/angles ($V_i$, $\theta_i$), and
$\bm{z}$ denotes the active/reactive power injections at buses ($P_i$, $Q_i$), $h(x)$ is the nonlinear equations vector correspondent to the  flows on branches ($P_{ij}$, $Q_{ij}$) that can be represented by AC power flow equations in~\eqref{eq:pf}:
\begin{equation} \label{eq:pf}
\begin{aligned}
P_i &= \sum_{j \in \Omega_i} V_i V_j (G_{ij}\cos\theta_{ij} + B_{ij}\sin\theta_{ij}) = {P_G}_i - {P_L}_i \\
Q_i &= \sum_{j \in \Omega_i} V_i V_j (G_{ij}\sin\theta_{ij} - B_{ij}\cos\theta_{ij}) = {Q_G}_i - {Q_L}_i \\
P_{ij} &= V_i^2(g_{si} + g_{ij}) - V_iV_j(g_{ij}\cos\theta_{ij} + b_{ij}\sin\theta_{ij}) \\[0.3em]
Q_{ij} &= -V_i^2(b_{si} + b_{ij}) - V_iV_j(g_{ij}\sin\theta_{ij} - b_{ij}\cos\theta_{ij}). \\
\end{aligned}
\end{equation}
where $\Omega_i$ represents the set of buses connected to bus $i$;
$G_{ij}+jB_{ij}$ corresponds to the $ij^{th}$ elements of bus admittance matrix $\bm{Y}$; and $g_{ij}+jb_{ij}$ denotes the series branch admittance between buses $i$ and $j$, respectively.

PSSE unit can be vulnerable to cyberattacks because if an adversary has `enough' knowledge about the power grid, s/he can shift the system state from its original value by injecting some false data to the measurements. In other words, if the adversary find vectors $\bm{a} \in \R^m$ and $\bm{c} \in \R^n$ that satisfy equation~\eqref{eq:fdia}, then they can easily add their attack vector $\bm{a}$ to the $\bm{z}$ and shift the state vector by $\bm{c}$ from its original value $\bm{x}$ without being detected by traditional BDD algorithms.
\begin{equation} \label{eq:fdia}
\bm{z} + \bm{a} = h({\bm{x + c}}),
\end{equation}


\section{Cyberattack detection by Chebyhsev Graph Convolutional Networks}\label{sec:methods}
\subsection{Chebyshev Graph Convolution}
Power system variables such as $\bm{P}$, $\bm{Q}$, $\bm{V}$, and $\bm{\theta} \in \R^n$ can be represented as graph signals by modeling the power grid as a graph.
Specifically, if we map buses to vertices $\V$ ($|\V| = n$), branches and transformers to edges ($\E$), and line admittances to weighted adjacency matrix ($\bm{W} \in \R^{n \times n}$), we can efficiently represent the power grid topology with a connected, undirected, weighted graph $\G = (\V, \E, \bm{W})$.
For $\G$, the normalized graph Laplacian can be given by~\eqref{eq:laplacian}.
\begin{equation}
\bm{L} = \bm{I_n} - \bm{D}^{-1/2} \bm{W} \bm{D}^{-1/2} \in \R^{n \times n}
\label{eq:laplacian}
\end{equation}
The Laplacian, i.e. $L$ is an essential operator in the theory of GSP.
Since $\bm{L}$ is a real, symmetric, and positive semi-definite matrix, it can be factorized as $\bm{L} = \bm{U} \bm{\Lambda} \bm{U}^T$
where $\bm{U} = [\bm{u}_0, \ldots, \bm{u}_{n-1}] \in \R^{n \times n}$ denotes the $n$ orthonormal eigenvectors, and $\bm{\Lambda} = \diag([\lambda_0, \ldots, \lambda_{n-1}]) \in \R^{n \times n}$ corresponds $n$ eigenvalues of $\G$.
Indeed, $\bm{u}_i$ and $\lambda_i$ values form the Fourier basis and Fourier frequencies in spectral domain of $\G$ \cite{ortega2018graph}.

Different from classical signal processing, there is no meaningful translation operation in the vertex domain \cite{gcnn}.
Thus, to filter a vertex domain graph signal $\bm{x} \in \R^{n}$ with a filter $g_\theta$ defined on $\G$, $\bm{x}$ is first transformed into the spectral domain using Graph Fourier Transformation (GFT) by $\tilde{\bm{x}} = \bm{U}^T \bm{x}$.
Next, the spectral domain signal $\tilde{\bm{x}} \in \R^{n}$ is filtered in the spectral domain by $\tilde{\bm{y}} =  g_\theta(\bm{\Lambda}
) \tilde{\bm{x}}$ where $g_\theta(\bm{\Lambda}) = \diag(\bm{\theta})$ is a filter kernel, and $\bm{\theta} \in \R^n$ is a vector of Fourier coefficients \cite{ortega2018graph}.
Finally, the result is transformed back to the vertex domain using inverse GFT by $\bm{y} = \bm{U}^T \tilde{\bm{y}}$ \cite{gcnn}.

\begin{figure*}[t]
	\centering
	\includegraphics[width=0.97\textwidth]{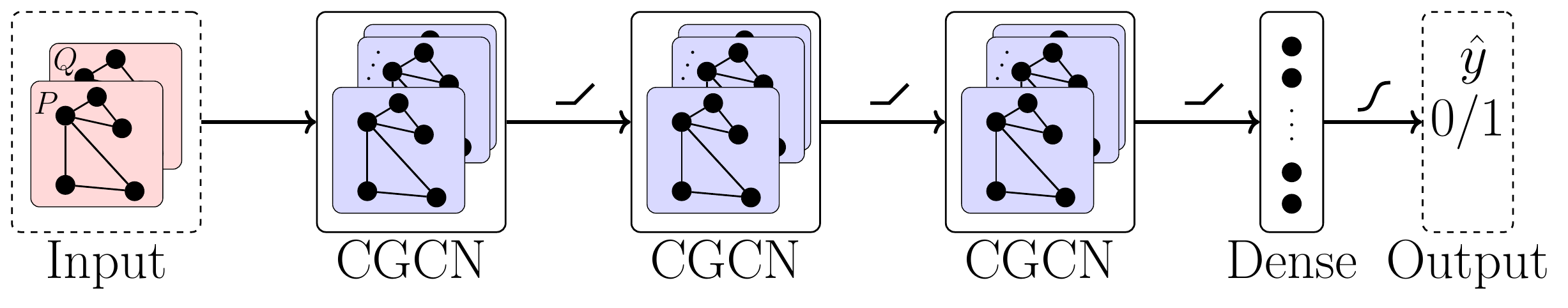}
	\caption{Architecture of the proposed GNN based detector. It takes active and reactive bus power injections $\bm{P}$ and $\bm{Q}$ as its inputs, extracts their spatial correlations in its Chebyshev GCN based hidden layers, weights the extracted features in its dense layer and produces the binary flag in its output layer. Note that while the input layer has two channels, the CGCN layers can have multiple channels.}
	\label{fig:arhitecture}
\end{figure*}

Although powerful, those spectral filters are computationally complex and spatially not localized due to the forward and inverse GFT operations.
To reduce their complexity and make them localized, Chebyshev convolutional graph filters are proposed in \cite{gcnn}.
Chebyshev polynomial of the first kind $T_k(x)$ with the order $K$ can be computed by a recursion as follows:	
\begin{equation} \label{eq:cheb_recur}
T_k(x) = 2x T_{k-1}(x) - T_{k-2}(x),
\end{equation}
where $T_0(x)=1$ and $T_1(x)=x$ \cite{mason2002chebyshev}.
Similarly, a filter $g_\theta$ can be computed by Chebyshev polynomial approximation, $T_k$, up to order $K-1$.
In this case, $g_\theta$ can filter $\bm{x}$ using the following equation:
\begin{equation} \label{eq:graph_conv_cheby}
\bm{y} = g_\theta \ast_\G \bm{x} = \sum_{k=0}^{K-1} \theta_k T_k(\tilde{L}) \bm{x},
\end{equation}
where $\bm{\theta} \in \R^K$ is a vector of Chebyshev coefficients, and $T_k(\tilde{\bm{L}}) \in \R^{n \times n}$ is the $K$ order Chebyshev polynomial evaluated at the scaled Laplacian $\tilde{L} = 2 L / \lambda_{max} - I_n$.
Slightly changing notation, $\bm{y}$ can be calculated by:
\begin{equation} \label{eq:final_res}
\bm{y} = \sum_{k=0}^{K-1} \theta_k \bar{\bm{x}}_k
\end{equation}
where $\bar{\bm{x}}_0=\bm{x}$, $\bar{\bm{x}}_1=\tilde{\bm{L}} \bm{x} $, and $\bar{\bm{x}}_k$ is computed recursively by:
\begin{equation} \label{eq:final_recur}
\bar{\bm{x}}_k = 2\tilde{\bm{L}}\bar{\bm{x}}_{k-1} - \bar{\bm{x}}_{k-2}.
\end{equation}

Note that Chebyshev polynomial approximation makes $g_\theta$ $K$-localized and reduces its computational complexity from $\bO(n^2)$ to $\bO(K|\E|)$. 
Therefore, Chebyshev Graph Convolutional operation can be effectively employed to capture the spatial correlations of the power grid data.
More details can be found in~\cite{gcnn, ortega2018graph}. 

\subsection{Architecture of the Proposed Detector}
The architecture of the proposed CGCN-based cyberattack detector is depicted in Fig.~\ref{fig:arhitecture} where the model inputs/outputs and hidden layers are outlined with dashed and solid blocks, respectively.
Due to the fact that $P_i+jQ_i = \sum_{k \in \Omega_i} P_{ik} +jQ_{ik}$, node values can be used to represent branch values as summation in their corresponding set of buses connected to them.
Therefore, we only employ $P_{i}$ and $Q_i$ values to feed the proposed model.
The model consists of $L$ hidden Chebyshev graph convolutional layers for spatial feature extraction and one dense layer for predicting the probability of the input sample being attacked.
In this multilayer architecture, the input, $X^{l-1} \in \R^{n \times c_{l-1}}$, and output, $X^{l} \in \R^{n \times c_{l}}$, of each CGCN layer are related by equation~\eqref{each CGCN layer }.
\begin{equation} \label{each CGCN layer }
\bm{X}^l = \textrm{ReLU}(\bm{\theta}^l \ast_\G \bm{X}^{l-1} + \bm{b}^l),
\end{equation}
where ReLU is rectified linear unit activation function, $\bm{\theta}^l \in \R^{K \times c_{l-1} \times c_{l}}$ is unknown trainable Chebyshev coefficients, $\bm{b}^l \in \R^{c_l}$ is bias term of the layer $l$, and $c_l$ is the number of channels in layer $l$ for $1 \leq l \leq L$.
Dense layer, on the contrary, gives $y$ in the classical feed-forward neural network fashion by:
\begin{equation}
y = \sigma(\bm{W}^L \bm{X}^L + \bm{b}^L)
\end{equation}
where $\bm{W}^L \in \R^{n \times c_L} $ is feature weights, $\bm{b}^L \in \R$ is the bias term and $\sigma$ is the nonlinear sigmoid activation operation.

\section{Numerical Experiments \& Discussions}\label{sec:results}
\subsection{Dataset Generation}
Due to the privacy reasons, there is no publicly available dataset for cyberattack detection, therefore, we generate a synthetic dataset.
As a first step, for each $t$  in $1 \leq  t \leq 36000$, we scale load and generation values of each bus in the 2848-bus test system \cite{josz2016ac} by a uniform random value between 0.8 and 1.2; run AC power flow algorithms \cite{pandapower}; and save power measurements after adding  1\% noise to them to mimic the timely behavior of the grid.
Then, to simulate the cyberattacks, we implement data scale attacks (\bm{$A_s$}) \cite{jevtic2018physics} and distribution-based attacks (\bm{$A_d$}) \cite{ozay2015machine} as of two frequently used cyberattack generation algorithms.
Scale attacks multiply the original measurement with a number sampled from a uniform distribution between 0.9 and 1.1.
In contrast, distribution-based attacks replace it with a value drawn from the Gaussian distribution satisfying the same mean and variance with the original measurements.

\subsection{Model Training}
We scale the dataset with the normal scaler for faster training, and split it into three sections to use 4/6 of them for training, 1/6 of them for validation, and 1/6 of them for testing.
The number of samples in each split is given in Table \ref{tab:sample} where we keep the number of attacked samples equal with the number of unattacked samples in each split for a balanced classification problem.
Each sample contains $P_{i}$, and $Q_i$ as input features and a binary output label $y$ to indicate the presence of the attack.
\begin{table}[h]
	\centering
	\caption{Number of samples in each split.}
	\setlength{\tabcolsep}{3pt}
	\renewcommand{\arraystretch}{1.2}
	\newcolumntype{?}[1]{!{\vrule width #1}}
	\begin{tabular}{c  c  c c  c}
		\textbf{split} & \textbf{non-attacked} & \textbf{$A_d$} & \textbf{$A_s$} & \textbf{total}  \\
		\specialrule{1pt}{1pt}{1pt}
		\textbf{train}      & 12000  & 6000 & 6000 & 24000 \\ \hline
		\textbf{validation} &  3000  & 1500 & 1500 &  6000 \\ \hline
		\textbf{test}       &  3000  & 1500 & 1500 &  6000 \\
	\end{tabular}
	\label{tab:sample}
\end{table}

 We utilize the binary cross-entropy loss function  in~\eqref{eqn:train} to compute all unknown parameters of the model represented with $ W_{\theta}$, by an end-to-end training process of $N$ training samples.
\begin{equation} \label{eqn:train}
L(\hat{y}, W_{\theta}) = \frac{-1}{N} \sum_{n=1}^{N}	y_i \log(\hat{y}_i) + (1-y_i)\log(1-\hat{y}_i),
\end{equation}
We feed samples into the model as mini batches having $2^8$ samples in $2^8$ maximum epoch. 
Moreover, we tolerate 16 epoch without any improvement in the validation set's cross entropy loss, otherwise we apply early stopping in order to avoid overfitting.
We run our implementations on Intel i9-8950 HK CPU \@ 2.90GHz with NVIDIA GeForce RTX 2070 GPU using Python 3.8 and Tensorflow 2.2 \cite{tensorflow}.

We also implement other existing deep learning based architectures in the literature such as FCN \cite{s97}, RRN \cite{s94}, and CNN \cite{lu2020false} to compare the proposed CGNN-based architecture with as we do not have access to the dataset of corresponding works.
For a fair comparison, we optimize the models' hyperparameters such as the number of hidden layers $\mathcal{L} = \{1,2,3,4,5\}$, the number of units $\mathcal{U} = \{8,16,32,64,128\}$, and the size of the filters $\mathcal{K} = \{3,5,7,9,\}$ using grid search.
Similar to the proposed model, we train the detectors on the training split and tune their hyperparameters on the validation split. Table \ref{tab:hpo} summarizes the optimized model hyperparameters for each model.
\begin{table}[h] 
	\centering
	\setlength{\tabcolsep}{3pt}
	\renewcommand{\arraystretch}{1.2}
	\newcolumntype{?}[1]{!{\vrule width #1}}
	\caption{Optimized model hyper-parameters.}
	\begin{tabular}{ccccc}
		\textbf{parameter} & \textbf{FCN} & \textbf{RNN} & \textbf{CNN} & \textbf{CGCN} \\
		\specialrule{1pt}{1pt}{1pt}
		$\mathcal{L}$ &  4 &  4 &  3 &  4 \\  \hline
		$\mathcal{U}$ & 64 & 32 & 32 & 32 \\  \hline
		$\mathcal{K}$ &  - &  - &  5 &  5 \\  
	\end{tabular}
	\label{tab:hpo}
\end{table}

\subsection{Detection Results}
\begin{figure}[h]
	\centering
	\includegraphics[width=0.47\textwidth]{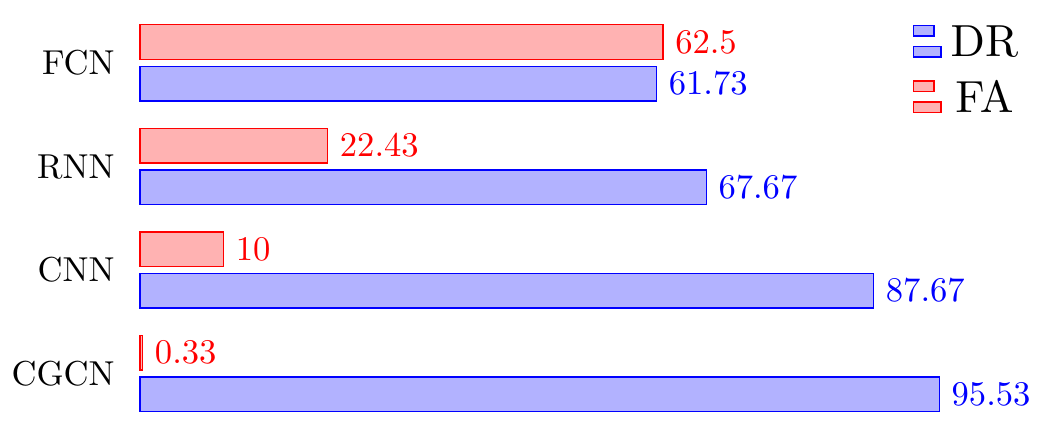}
	\caption{Detection results in terms of $DR$ and $FA$ percentages.}
	\label{fig:results}
\end{figure}

We evaluate the model performance by the detection rate $DR = \frac{TP}{TP+FN}$ and false alarm rate $FA = \frac{FP}{FP + TN}$ where $TP$, $FP$, $TN$, and $FN$ denote true positives, false positives, true negatives, and false negatives, respectively.
Fig~\ref{fig:results} presents the detection results of each model for the 2848 bus test system. 
Clearly, FCN falls behind other models since it has the lowest $DR$ with $55.06\%$ and highest $FA$ with $62.5\%$.
RNN performs better than FCN with $71.19\%$ $DR$ and $22.43\%$ $DR$.
Compared to the non-convolutional architectures, i.e., FCN and RNN, the convolutional architectures, i.e., CNN and CGCN, give better results. 
Additionally, CGNN surpasses CNN by 7.86\% in $DR$ and 9.67\% in $FA$.

Our experiments indicate the importance of architectural choices on the models' detection performances.
For instance, FCN falls short since it ignores the locality of features and overfits to the training data.
RNN, on the contrary, does not yield convenient results due to the fact that structure of power grid data do not fit well into sequence type recurrent relations.
Compared to FCN and RNN, CNN gives better results since it can capture the temporal or spatial correlations of the input data in an Euclidean space in which local features can be expressed linearly.
Except some trivial grids, the spatial relationship of the power grid measurements can not be fully modeled in an Euclidean space due to the its graph structure.
In fact, power grid data needs topology-aware architectures such as CGCN to better exploit its spatial correlations.

\subsection{Detection Times} \label{sec:time}
Detection time of a detector can be highly critical in a practical application since PSSE outcome is directly fed into various time sensitive Energy  Management System (EMS) blocks including  contingency and reliability analysis, load and price forecasting, and economic dispatch processes \cite{abur2004power}.
To compare the detection time of different algorithms, we measure the elapsed time during the model's detection process for each sample in the test set and tabulate the mean values in Table~\ref{tab:time}.	
\begin{table}[h!] 
	\centering
	\caption{Models' detection times in milliseconds.}
	\setlength{\tabcolsep}{3pt}
	\renewcommand{\arraystretch}{1.2}
	\newcolumntype{?}[1]{!{\vrule width #1}}
	\begin{tabular}{c c c c}
		\textbf{FCN} & \textbf{RNN} & \textbf{CNN} & \textbf{CGCN} \\ \specialrule{1pt}{1pt}{1pt}
		1.33  & 1125.66 & 3.54 & 3.25 \\
	\end{tabular}
	\label{tab:time}
\end{table}

As can be seen from Table~\ref{tab:time}, RNN's detection time is not acceptable for real time application since it takes more than a second to respond.
FCN, in contrast, yields the best detection delay with only 1.33 ms. 
Yet, its unsatisfactory $DR$ and $FA$ confine its suitability for real time application.
CNN and CGCN give acceptable delays for a practical scenario with 3.54 and 3.25 ms detection times, respectively.
Besides, CGCN provides better results in terms of detection performance and delay.

\section{Conclusion} \label{sec:conclusion}
Modern power grids are vulnerable to cyberattacks due to their highly complex and integrated cyber-physical networks.
Although a number of solutions have been proposed to detect those cyberattacks, 
most of the studies have disregarded the inherent topology of the power grid and used small test systems to verify their algorithms.
To address these issues and detect cyberattacks in large scale AC power grids, we propose a deep learning model that 
employs Graph Convolutional Networks in its hidden layers to better capture power grid measurements' spatial correlations.
It is numerically verified on a large-scale power grid with 2848 buses that the proposed detector outperforms state-of-the-art model by 7.86\% and 9.67\% in false alarm rate and detection rate, respectively.

\section*{Acknowledgment}
This work was supported by NSF under Award Number 1808064.

\bibliographystyle{IEEEtran}
\bibliography{iceee2022}

%
%
%

\end{document}